\documentclass[11pt,english]{article}

\usepackage[maccyr]{inputenc}

\usepackage{mathrsfs}

\usepackage[margin=2cm]{geometry}

\usepackage{amssymb,amsmath}
\usepackage{cite}

 \usepackage[colorlinks,citecolor=blue,urlcolor=blue,bookmarks=false,hypertexnames=true]{hyperref}

\RequirePackage{amsmath,amssymb,amsxtra,amsthm}

\RequirePackage{mathrsfs}

\RequirePackage{setspace}
\RequirePackage{array}
\RequirePackage{booktabs}

\RequirePackage{braket}

\RequirePackage[pdftex,final]{graphicx}
\graphicspath{{Plots/}}

\RequirePackage{colortbl}
\RequirePackage{xcolor}
\colorlet{titlerowcolor}{gray!15}

\usepackage{cite}

\numberwithin{equation}{section}
\numberwithin{table}{section}
\numberwithin{figure}{section}

\author{
  \begin{minipage}{1.00\linewidth}
    \vspace{0.5cm}
    \begin{center}
      \begin{small}
        \ \ \textbf{Carlo Angelantonj} $^{1,2}$, \textbf{Ignatios Antoniadis} $^{3,4,5}$, \textbf{Ioannis Florakis} $^{6}$ and \textbf{Hongliang Jiang} $^{7}$
     \end{small}
    \end{center}
    \vspace{.3cm} \hspace{0.75cm}\begin{minipage}{.85\linewidth}
      {\it \begin{footnotesize}
          \begin{itemize}
         \item[${}^1$] Dipartimento di Fisica, Universit\`a di Torino, and INFN Sezione di Torino
          \\
            Via Pietro Giuria 1, 10125 Torino, Italy
         \item[${}^2$] Arnold-Regge Center, Via Pietro Giuria 1, 10125 Torino, Italy 
         \item[${}^3$] Laboratoire de Physique Th\'eorique et Hautes \'Energies - LPTHE\\
Sorbonne Universit\'e, CNRS, 4 Place Jussieu, 75005 Paris, France
	\item[${}^4$] Nordita,
Stockholm University and KTH Royal Institute of Technology\\
Hannes Alfv\'ens v\"ag 12, 106 91 Stockholm, Sweden
	\item[${}^5$] Department of Mathematical Sciences, University of Liverpool, 
Liverpool L69 7ZL, United Kingdom
	\item[${}^6$] Department of Physics, University of Ioannina
	\\
	45110 Ioannina, Greece
	\item[${}^7$] Centre for Theoretical Physics, Department of Physics and Astronomy 
	\\
	Queen Mary University of London, Mile End Road, E1 4NS, UK
\end{itemize}
        \end{footnotesize}}
    \end{minipage}
    \vspace{1cm}
  \end{minipage}
}

\date{}

\title{\vspace{0.5cm}
  \begin{huge}
    \textbf{Refined Topological Amplitudes from the\\[15pt] $\mathbf\Omega$-Background in String Theory} 
  \end{huge}
}

\begin{document}

\begin{titlepage}
  \maketitle
  \thispagestyle{empty}

  \vspace{-14cm}
  \begin{flushright}
 QMUL-PH-22-07
   \end{flushright}

  \vspace{11cm}

  \begin{center}
    \textsc{Abstract}\\
  \end{center}

It was recently shown that the ${\cal N}=2$ string topological amplitudes in the heterotic weak coupling limit generate a six-dimensional Melvin space, providing a description of the $\Omega$-background in string theory, where string propagation can be exactly studied. In this work, we generalise the analysis to the refined case of the $\Omega$-background with two independent deformation parameters. The Melvin space is now ten-dimensional and is extended by an action on the internal K3 compactification manifold of the heterotic superstring, corresponding to an ${\rm SU}(2)_R$ rotation in the field theory description. We identify the class of heterotic topological amplitudes realising this background as the scattering of two anti-self-dual gravitons and arbitrary numbers of anti-self-dual graviphotons, self-dual vector fields of the dilaton multiplet, together with self-dual magnetic fluxes along the K3. In the field theory limit, our result correctly reproduces the perturbative part of the Nekrasov free energy in the case where both equivariant parameters are turned on.

\vfill

\begin{center}
	\emph{Dedicated to the memory of Costas Kounnas and Theodore Tomaras.}
\end{center}

\vfill

{\small
\begin{itemize}
\item[E-mail:] {\tt carlo.angelantonj@unito.it}
\\
{\tt ignatios.antoniadis@upmc.fr}
\\
{\tt iflorakis@uoi.gr}
\\
{\tt h.jiang@qmul.ac.uk}
\end{itemize}
}
\vfill

\end{titlepage}




\section{Introduction}

An interesting open problem is to unveil the string geometry behind topological amplitudes~\cite{Angelantonj:2019qfw}. In the simplest case, the genus-$g$ partition function $\mathcal{F}_g$ of the topological string~\cite{Bershadsky:1993cx}, obtained by an appropriate twist of the $\mathcal{N}=2$ superconformal $\sigma$-model~\cite{Witten:1988xj, Witten:1989ig} describing the type II compactification on a 6d Calabi-Yau (CY) manifold, computes the coefficients of the higher-derivative $F$-term $W^{2g}$ in the string effective action \cite{Antoniadis:1993ze}, and their dependence on vector multiplet moduli. The (chiral) Weyl superfield  $W=F_{(-)}^G+\theta R_{(-)}\theta+\dots$~\cite{deRoo:1980mm, Bergshoeff:1980is}, with $\theta$ denoting the $\mathcal{N}=2$ fermionic coordinates, contains the anti-self-dual graviphoton field strength as lowest component, along with the anti-self-dual Riemann tensor. As a result, the $\mathcal{F}_g$'s can be extracted from a physical amplitude involving two (anti-self-dual) Riemann tensors and $2g-2$ (anti-self-dual) graviphoton field strengths. It was later observed \cite{Nekrasov:2002qd,Nekrasov:2003rj} that, in an appropriate field theory limit around the conifold point and upon identifying the Weyl superfield $W$ with $\hbar$, the higher-derivative couplings ${\cal F}\equiv\sum_g \mathcal{F}_g W^{2g}$ reproduce the instanton-corrected partition function of $\mathcal{N}=2$ supersymmetric Yang-Mills theory, regularised by the $\Omega$-background when a single equivariant parameter $\hbar=\epsilon_{-}$ is turned on \cite{Moore:1997dj, Losev:1997wp, Nekrasov:2002qd, Nekrasov:2003rj}.

Using heterotic/type IIA string duality, all $\mathcal{F}_g$'s can be studied in the weak coupling limit on the heterotic side at the one loop level, after the string dilaton is properly mapped to a particular $\mathcal{N}=2$ vector multiplet associated to the base modulus of CY manifolds which are K3 fibrations~\cite{Antoniadis:1995zn}. This has the advantage of clarifying the connection with the dynamics of $\mathcal{N}=2$ supersymmetric Yang-Mills in an appropriate field theory limit, where the heterotic couplings ${\cal F}_{\rm het}$ reproduce the perturbative part of the Nekrasov partition function.

In recent work~\cite{Angelantonj:2019qfw}, it was shown that the generating function of the topological string amplitudes of \cite{Antoniadis:1995zn} generate a geometric background, corresponding to a six-dimensional (6d) Melvin spacetime \cite{Melvin:1963qx}, amenable to an exact (super)conformal field theory description~\cite{Russo:1994cv,Russo:1995tj, Russo:1995aj}. More precisely, it was shown that, in the heterotic weak coupling limit, the F-term series ${\cal F}_{\rm het}$ identically reproduces  the heterotic one-loop vacuum energy on the 6d Melvin background. This equivalence provides strong evidence that the Melvin spacetime is indeed the correct string realisation of the $\Omega$-background, in agreement with  \cite{Hellerman:2011mv, Hellerman:2012zf, Orlando:2013yea, Lambert:2014fma}. Indeed, the 6d (Euclidean) Melvin space is flat and can be realised as a freely-acting orbifold where equal and opposite rotations on two planes are accompanied by a winding around a cycle inside a two-torus $T^2$. The rotation angles are clearly identified with the deformation parameters of the $\Omega$-background. Even though this background spontaneously breaks half of the supersymmetries, the partition function does not vanish, since the whole theory can be viewed as two-dimensional (in the non-compact limit of $T^2$), with $(4,0)$ supersymmetry, which does not necessarily imply an equality of bosonic and fermionic degrees of freedom.

The matching between topological amplitudes and the Nekrasov free energy has been conjectured to hold also when the second equivariant parameter $\epsilon_{+}$ of the $\Omega$ background is turned on \cite{Nekrasov:2002qd, Nekrasov:2003rj}. It is then desirable to ``refine" the topological amplitudes of \cite{Antoniadis:1993ze,Antoniadis:1995zn}, to account for this extra deformation.
Indeed, from the gauge theory perspective, the full $\Omega$-background involves two independent angles $\epsilon_{1,2}$ rotating the two planes of Euclidean spacetime, properly accompanied by an ${\rm SU}(2)_R$ rotation by the angle $\epsilon_+=\epsilon_1+\epsilon_2$ in order to preserve supersymmetry. 
A concrete realisation of this refinement, involving new topological amplitudes, turns out to be a highly non-trivial task.
On the one hand, a non-vanishing $\epsilon_+$ calls for additional vertex operator insertions associated to self-dual gauge field strengths, which typically spoil the topological nature of the amplitudes, since string oscillators no longer decouple and (quasi-)holomorphicity is lost. On the other hand, in the type II description and aside from the $\mathcal{N}=2$ graviphoton, all vector multiplets are rotated into each other by duality transformations and, a priori, there appears to be no natural choice for the extra vector.

In the last couple of decades, several attempts have been made to pinpoint the precise vertex operators realising the refinement.
For instance, in \cite{Antoniadis:2010iq} the additional insertions were chosen to be the self-dual field strengths of the vector multiplet containing the heterotic dilaton. Although this choice does yield topological amplitudes, it fails to correctly reproduce the perturbative part of the Nekrasov partition function in the field theory limit.
In a later work  \cite{Nakayama:2011be}, insertions of field strengths of the vector partners of the K\"ahler and complex structure moduli of the internal $T^2$, as well as of the U(1) current of the superconformal algebra were considered. However, the resulting amplitudes do not correspond to a Gaussian $\sigma$-model, but involve higher-order corrections, preventing an exact evaluation at the full string level.
Subsequently, in \cite{Antoniadis:2013bja} the vector field strengths in the dilaton multiplet were replaced by those in the multiplet of the K\"ahler modulus of $T^2$, or equivalently by the D5-brane gauge coupling on the type I side. In this case, although the amplitudes reproduce the correct field theory limit, even at the non-perturbative level \cite{Antoniadis:2013mna}, their topological nature is lost and the would-be holomorphic anomaly equation mixes them with new terms that obstruct their interpretation as the free energy of the refined topological string. Moreover, the decoupling of hypermultiplets is not manifest at the full string theory level.

In this work, we identify the refinement of the topological amplitudes, in the heterotic weak coupling limit, corresponding to the full $\Omega$-background. This is accomplished by implementing the program initiated in \cite{Angelantonj:2019qfw}, whereby the exponentiation of topological amplitudes is conjectured to correspond to solvable backgrounds in string theory. As a starting point, we geometrically engineer the second equivariant parameter $\epsilon_{+}$ associated to the R-symmetry twist, as a further Melvin rotation of the K3 coordinates, now accompanying the independent rotations $\epsilon_{1,2}$ along the two planes of Euclidean spacetime.
This background is now eight-dimensional, and preserves 1/8 of the original 10d supersymmetries, provided that the sum of angles vanishes, \emph{i.e.} $\epsilon_{+}=\epsilon_1+\epsilon_2$. In the decompactification limit of $T^2$, the resulting theory can be viewed as two-dimensional with $\mathcal N=(2,0)$ supersymmetry, whose representations are again no longer required to be Bose-Fermi degenerate. We explicitly evaluate the partition function of the heterotic string on this 8d Melvin space and, in the field theory limit, we obtain a precise matching with the perturbative part of the $\mathcal{N}=2$ gauge theory free energy of Nekrasov when both parameters are turned on. This match provides strong evidence that our generalised Melvin space yields the correct string uplift of the $\Omega$-background.

Rotations in Melvin spaces can be alternatively viewed as magnetic flux backgrounds which, in our setup, lie both in spacetime and along the internal K3 surface. This observation, together with the geometric interpretation of the exponentiated topological amplitudes \cite{Angelantonj:2019qfw}, allows us to identify the new vertex operators of the self-dual fields realising the refinement.
We show that besides the anti-self-dual graviphoton field strengths, the additional vertex operators entering the amplitudes are associated with the self-dual field strengths of the heterotic dilaton vector multiplet, as in~\cite{Antoniadis:2010iq}, together with an appropriate number of self-dual magnetic fluxes along the K3 manifold, emanating from the R-symmetry twist. Notice that, although the cohomology $H^1({\rm K3})$ vanishes and, therefore, the components of the 10d gauge fields along the K3 directions are absent, their fluxes are still allowed. For instance, in the orbifold limit of ${\rm K3}\sim T^4/\mathbb{Z}_N$, they are invariant under the orbifold action, even though the associated gauge potential components are not. Although the compactness of K3 imposes a quantisation on the flux and thus on the deformation parameter $\epsilon_+$, the final result may be analytically continued to any value.
The refined amplitudes can then be exponentiated into a Gaussian $\sigma$-model, where both spacetime and internal string coordinates are holomorphically rotated by the independent $\epsilon$ parameters. As expected, the functional integral precisely reproduces the heterotic string partition function on our Melvin background and, therefore, fully matches the perturbative free energy of Nekrasov, in the field theory limit.

The outline of the paper is as follows. In Section~\ref{Section2}, we describe the K3 compactification of the heterotic string on the $\Omega$-background represented by the supersymmetric Melvin space involving rotations on $\mathbb R^4\times {\rm K3}$ coupled to shifts along the $T^2$, in the orbifold K3 limit. We explicitly compute the one-loop partition function and verify the complete decoupling of hypermultiplet moduli. Expanded around a point of SU(2) gauge symmetry enhancement, it reproduces the perturbative part of the Nekrasov partition function, in the field theory limit. We further show how the same analysis may be extended to the case of $\mathcal{N}=2^\star$ gauge theories.
In Section~\ref{Section3}, utilising the Melvin geometry, we deduce the deformed $\sigma$-model leading to the generating function of the refined topological amplitudes and identify the new vertex operators involved in our proposal. We compute the functional integral at genus one and show that the result matches precisely the partition function on the generalised Melvin background of Section~\ref{Section2}. In Section~\ref{Section4}, we identify the higher-derivative F-terms computed by the refined topological amplitudes and relate their coefficients $\mathcal{F}_{g,n,m}$ to those entering the $\epsilon_{\pm}$-expansion of the Nekrasov free energy. Finally, Section~\ref{SectionConclusions} contains our concluding remarks.


\section{$\mathcal{N}=2$ Heterotic String on the $\Omega$-Background} \label{Section2}

In heterotic string theory, 4d vacua with $\mathcal{N}=2$ supersymmetry naturally emerge from the compactification on $\text{K3}\times T^2$ spaces. In the orbifold limit $\text{K3} = T^4 /\mathbb{Z}_N$, $N=2,3,4,6$,  these vacua admit a fully-fledged worldsheet description. Without loss of generality, we focus on the  case $N=2$, whereby the complex coordinates $(z^3, z^4)$ along the singular $\text{K3}$ undergo a rotation by an angle $\theta=\pi$. Modular invariance requires a non-trivial action on the gauge bundle. For simplicity, we identify the gauge connection with the spin connection via the standard embedding. Of course, other choices of the orbifold group $\mathbb{Z}_N$ and different choices of gauge bundle are possible, but will not affect our conclusions. The partition function for the $\text{E}_8 \times \text{E}_8$ heterotic string reads
\begin{equation}
	{\mathcal Z} =  \tfrac{1}{2} \sum_{h,g=0,1} \tfrac{1}{2}\, \sum_{a,b=0,1} (-1)^{a+b+ab}\, 
\frac{\theta \big[ {\textstyle{a \atop b}}\big]^2 \,\theta \big[ {\textstyle{a +h\atop b+g}}\big]\,\theta\big[ {\textstyle{a -h\atop b-g}}\big]}{\eta^6 \,\bar\eta^2} \,
\Gamma_{4,4} \big[ {\textstyle{h \atop g}}\big]\, \Gamma_{2,2} (T,U)\, \bar \Gamma_8 \big[ {\textstyle{h\atop g}}\big]\, \bar \Gamma_8 \big[ {\textstyle{0\atop 0}}\big] \,.
\label{ZN2}
\end{equation}
Here  $\eta$ is the Dedekind function, $\theta \big[ {\textstyle{\alpha \atop \beta}}\big]$ is the Jacobi theta constant with characteristics\footnote{In this paper we adopt the conventions of \cite{Kiritsis:2019npv} for theta functions with characteristics. Notice that these differ from those used by some of the authors in other works, for instance \cite{Angelantonj:2019qfw}.} $(a,b)$, while $h$ labels the $\mathbb{Z}_2$ (un)twisted sectors, and the sum over $g$ implements the orbifold projection. The sum over $a,b$ takes into account the various spin structures with a conventional choice of the GSO phases. $\Gamma_{4,4}$ encodes the contribution of the $Z^3$ and $Z^4$ bosonic coordinates along the $T^4 /\mathbb{Z}_2$. We employ a notation according to which the lower case $z^i$ denote the complex Cartesian coordinates of K3, whereas upper case $Z^i$ are the worldsheet scalars associated to the target space embedding, and similarly for the remaining spacetime and $T^2$ directions. The (0,0) sector involves the (4,4) Narain lattice divided by the appropriate Dedekind functions, while
\begin{equation}
	\Gamma_{4,4} \big[ {\textstyle{0\atop 1}}\big] = 16 \, \frac{\eta^2 }{\theta \big[ {\textstyle{1 \atop 0}}\big]^2} \, \frac{\bar\eta^2 }{\bar\theta \big[ {\textstyle{1 \atop 0}}\big]^2} \,,
\end{equation}
is sufficient to specify also the remaining twisted sectors, by means of modular invariance. $\Gamma_{2,2}$ involves the Narain lattice for the spectator $T^2$ with complex structure $U$ and K\"ahler modulus $T$. Finally, the first $\bar \Gamma_8$ with $(h,g)$ characteristics describes the broken  $\text{E}_8 \to \text{SU} (2) \times \text{E}_7$ factor, while the second $\bar \Gamma_8$ refers to the unbroken $\text{E}_8$.  Actually, this vacuum can be deformed by introducing Wilson lines $Y$ along the $(2,2)$ lattice, which would result in further breaking of the gauge group. For our purposes, we turn on Wilson lines in the unbroken $\text{E}_8$, so that
\begin{equation}
 \Gamma_{2,2} (T,U) \, \bar \Gamma_8 \big[ {\textstyle{0\atop 0}}\big] \to \Gamma_{2,10} (T,U,Y)\,.
\end{equation}

Having specified the $\mathcal{N}=2$ vacuum on a flat Euclidean space-time, we are now ready to implement the $\Omega$ deformation. Following \cite{Angelantonj:2019qfw}, we describe the latter as a Melvin background \cite{Russo:1994cv,Russo:1995tj,Russo:1995aj}, whereby the Euclidean space-time, with complex coordinates $(z^1, z^2)$, is fibered over the $T^2$ via the introduction of the non-trivial monodromy 
\begin{equation}
z^i  \to  e^{2 i \pi \epsilon_i  (\tilde m_1 + U_1 \tilde m_2 )}\, z^i  \qquad \text{as}\qquad x \to x + 2\pi \sqrt{\frac{T_2}{U_2}}\left( \tilde m_1 + U \tilde m_2\right) \,,
\end{equation}
where $i=1,2$, $x$ is the complex coordinate on the $T^2$ and $\tilde m_1 , \tilde m_2 \in \mathbb{Z}$ identify a generic one-cycle $\tilde m_1 + U \tilde m_2$ in $T^2$.

Actually, this deformation with arbitrary $\epsilon_1$ and $\epsilon_2$ breaks supersymmetry unless it is supplemented by an additional $\epsilon_+ = \epsilon_1 + \epsilon_2$ rotation along the $\text{SU} (2)_R$ $R$-symmetry. From a higher-dimensional perspective, the $R$-symmetry group emerges from properties of the internal CY space and, in the case at hand, it is given by an $\text{SU} (2)$ subgroup of the $\text{SO} (4)$ diffeomorphisms of the $T^4$. Note, however, that compact CYs do not admit continuous isometries, which reflects the fact that in (quantum) supergravity $R$-symmetry is either gauged or broken to a discrete subgroup. As a result, the Melvin deformation associated to the $\text{SU} (2)_R$ rotation can only be realised for rational values of $\epsilon_+$. Although we take this into account in our analysis, the final expressions can be analytically continued to arbitrary values. Our choice of $R$-symmetry embedding in $\text{SO} (4)$ implies the additional monodromies on the $(z^3,z^4)$ coordinates
\begin{equation}
z^{3,4} \to e^{ - i \pi \epsilon_+  (\tilde m_1 + U_1 \tilde m_2 )}\, z^{3,4}\,.
\end{equation}
All in all, this Melvin deformation is equivalent to a freely acting orbifold, and supersymmetry is preserved provided the sum of the rotation angles vanishes. In our case, this translates to the constraint $\epsilon_1 + \epsilon_2 - \epsilon_+  =0$.

The combination of the $\Omega$ deformation with the $\mathbb{Z}_2$ orbifold action is then equivalent to the following (freely-acting) rotations of the string fields associated to the $\mathbb{C}^2\times T^4$ fibre 
\begin{equation}
	\begin{split}
		Z^1(\sigma_1+\alpha,\sigma_2+\beta) &= e^{2\pi i \epsilon_1 \left[ (\tilde m_1+U_1\tilde m_2)\alpha+(n_1+U_1 n_2)\beta\right]} \, Z^1(\sigma_1,\sigma_2) \,,
		\\
		Z^2(\sigma_1+\alpha,\sigma_2+\beta) &= e^{2\pi i \epsilon_2 \left[ (\tilde m_1+U_1\tilde m_2)\alpha+(n_1+U_1 n_2)\beta\right]} \, Z^2(\sigma_1,\sigma_2) \,,
		\\
		Z^3(\sigma_1+\alpha,\sigma_2+\beta) &= e^{-\pi i \epsilon_+ \left[ (\tilde m_1+U_1\tilde m_2)\alpha+(n_1+U_1 n_2)\beta\right]+\pi i(g \alpha+h\beta)} \, Z^3(\sigma_1,\sigma_2)  \,,
		\\
		Z^4(\sigma_1+\alpha,\sigma_2+\beta) &= e^{-\pi i \epsilon_+ \left[ (\tilde m_1+U_1\tilde m_2)\alpha+(n_1+U_1 n_2)\beta\right]-\pi i(g \alpha+h\beta)} \, Z^4(\sigma_1,\sigma_2) \,,
	\end{split}
\end{equation}
coupled to the shift in the $T^2$ base
\begin{equation}
 	X(\sigma_1+\alpha,\sigma_2+\beta) = X(\sigma_1,\sigma_2) + 2\pi  \sqrt{\frac{T_2}{U_2}}\left[ ( \tilde m_1 + U \tilde m_2)\alpha+ (n_1+Un_2)\beta \right] \,,
\end{equation}
where $n_1, n_2 \in \mathbb{Z}$, and $\alpha,\beta=0,1$ define the boundary conditions along the two one-cycles of the worldsheet torus of complex structure $\tau=\tau_1+i\tau_2$.

Together with worldsheet supersymmetry, the above boundary conditions imply similar monodromies for the worldsheet fermions, and are sufficient to determine the partition function, which now reads\footnote{In the following we use the convention $\vartheta_1(z)=\theta\big[ {\textstyle{ 1\atop 1}} \big](z)$.}
\begin{equation}
\begin{split}
			\mathcal{Z} &= \tfrac{1}{2}{\sum_{h,g=0,1}} {\sum_{\tilde m,n,Q}}'  \left| \frac{\eta^2}{\vartheta_1(\chi_1)\,\vartheta_1(\chi_2)}\right|^2 e^{-\frac{\pi}{2\tau_2}(\chi_1-\bar\chi_1)^2-\frac{\pi}{2\tau_2}(\chi_2-\bar\chi_2)^2} 
			\\
		     & \times \tfrac{1}{2}\sum_{a,b=0,1}(-1)^{a+b+ab} \,\frac{\theta\big[ {\textstyle{ a\atop b}} \big](\chi_1)\,\theta\big[ {\textstyle{ a\atop b}} \big](\chi_2)\,\theta\big[ {\textstyle{ a+h\atop b+g}} \big](\frac{\chi_{+}}{2})\, \theta\big[ {\textstyle{ a-h\atop b-g}} \big](\frac{\chi_{+}}{2}) }{\eta^4} ~ e^{\frac{\pi}{2\tau_2}\left( \chi_1^2+\chi_2^2+\frac{1}{2}\chi_{+}^2-2|\chi_1|^2-2|\chi_2|^2-|\chi_{+}|^2\right)} 
		     \\
		     & \times \zeta\big[ {\textstyle{ h\atop g}} \big](\tfrac{\epsilon_{+}}{2})\left|\frac{\eta^2}{\theta\big[ {\textstyle{ 1+h\atop 1+g}} \big](\frac{\chi_{+}}{2})\, \theta\big[ {\textstyle{ 1-h\atop 1-g}} \big](\frac{\chi_{+}}{2})} \right|^2 ~ e^{-\frac{\pi}{4\tau_2}(\chi_{+}-\bar\chi_{+})^2} 
		     \\
	             & \times  [\Gamma_{2,10}(T,U,Y)]_{\tilde m,n,Q} \, \bar \Gamma_8 \big[ {\textstyle{ h\atop g}} \big] \,.
\end{split}
\label{HetMelK3}
\end{equation}
In this expression, $\chi_i = \epsilon_i \left(\tilde m_1+U_1\tilde m_2+\tau(n_1+U_1 n_2)\right)$, where $i=1,2,+$, and the vector $Q$ spans the $\text{E}_8$ charge lattice. The contribution from $(\tilde m,n)=(0,0)$ effectively sets $\chi_i\to 0$ and corresponds to the standard K3 compactification. In this case, supersymmetry implies that the partition function vanishes identically for any $h$ and $g$, as can be readily seen from Jacobi's \emph{aequatio identica satis abstrusa}. This is reflected in \eqref{HetMelK3} by the appearance of the primed sum, which excludes the term $({\tilde m},n)=(0,0)$. As a result, there is no dependence on the $\Gamma_{4,4}$ moduli (which would appear only in the purely untwisted $h=g=0$ sector for ${\tilde m}=n=0$) and, thus, the hypermultiplet contribution decouples.
 The first line is the contribution of the worldsheet bosons in the Euclidean $\mathbb C^2$ space, the second line is the contribution of the worldsheet fermions, the third line is the contribution of the worldsheet bosons along $T^4/\mathbb{Z}_2$, and finally the fourth line is the contribution of the gauge degrees of freedom together with the worldsheet bosons parametrising $T^2$. Note that, $[\Gamma_{2,10}(T,U,Y)]_{\tilde m,n,Q}$ is in the Lagrangian representation, where $\tilde m$ denotes the wrapping numbers along the first cycle of $T^2$ which, upon Poisson resummation, would become the Kaluza-Klein momenta. However, the dependence of the theta functions on $\tilde m$ makes this Poisson resummation rather cumbersome. In the untwisted $h=0, n=0$ sector,  
\begin{equation}
	\zeta\big[ {\textstyle{ 0\atop g}} \big](\tfrac{\epsilon_{+}}{2}) = 16 \sin^2\left[\tfrac{\pi}{2} (\epsilon_{+}(\tilde m_1+U_1\tilde m_2) + g) \right] \sin^2\left[\tfrac{\pi}{2}(\epsilon_{+} (\tilde m_1+U_1\tilde m_2)- g) \right]
\end{equation}
counts the $Z^3, Z^4$ zero modes, while modular invariance uniquely determines the remaining sectors.

The sum over spin structures can be performed using the generalised Jacobi identity
\begin{equation}
	\tfrac{1}{2}\sum_{a,b=0,1}(-1)^{a+b+ab} \prod_{k=1}^4 \theta\big[ {\textstyle{ a+h_k\atop b+g_k}} \big](z_k) = - \prod_{k=1}^{4} \theta\big[ {\textstyle{ 1+\tilde h_k\atop 1+\tilde g_k}} \big](\tilde z_k)\,,
\end{equation}
where 
\begin{equation}
	\begin{pmatrix}
		\tilde h_1 \\
		\tilde h_2 \\
		\tilde h_3 \\
		\tilde h_4 \\
	\end{pmatrix} = \tfrac{1}{2} \begin{pmatrix}
							1 & 1 & 1 & -1 \\ 1 & 1 & -1 & 1 \\ 1 & -1 & 1 & 1 \\ -1 & 1 & 1 & 1 \end{pmatrix} 	\begin{pmatrix}
		h_1 \\
		 h_2 \\
		 h_3 \\
		 h_4 \\
	\end{pmatrix} \,, 
\end{equation}
with similar expressions for $\tilde g_k$ and $\tilde z_k$. As a result, the worldsheet fermions exactly cancel the oscillators from the left-moving worldsheet bosons, so that only BPS states contribute to the partition function, which now reads
\begin{equation}
		\mathcal {Z} = -\tfrac{1}{2} {\sum_{h,g=0,1}} {\sum_{\tilde m,n,Q}}'  \,\frac{\zeta\big[ {\textstyle{ h\atop g}} \big](\tfrac{\epsilon_{+}}{2})\,e^{-\frac{\pi}{2\tau_2}(\bar\chi_1^2+\bar\chi_2^2+\frac{1}{2}\bar\chi_{+}^2)} }{ \bar\eta^{4}\, \bar\vartheta_1(\chi_1)\, \bar\vartheta_1(\chi_2)\,\bar\theta\big[ {\textstyle{ 1+h\atop 1+g}} \big](\frac{\chi_{+}}{2})\, \bar\theta\big[ {\textstyle{ 1-h\atop 1-g}} \big](\frac{\chi_{+}}{2})  } \,
		  \,[\Lambda_{2,10}(T,U,Y)]_{\tilde m,n,Q} \, \bar \Gamma_8 \big[ {\textstyle{ h\atop g}} \big] \,.
\label{BPSfree}
\end{equation}
In the above, we have introduced the Narain lattice $\Lambda_{2,10} = \eta^2\,\bar\eta^{10}\,\Gamma_{2,10}$, which indeed includes contributions only from the zero modes of the $(2,10)$ currents.

To make contact with the perturbative part of Nekrasov's partition function, we need to extract the field theory limit of \eqref{BPSfree} for an appropriate choice of gauge group, for instance, $\text{SU}(2)$. This is achieved by sending $\alpha'\to 0$, so that massive string modes and $n$-winding states are suppressed, while keeping the distance from the SU(2) enhancement point fixed and the dominant contribution arises from the $h=0$ untwisted sector. In this limit, gravity decouples and R-symmetry is no longer quantised. Therefore, even though in eq. \eqref{BPSfree} $\epsilon_{+}$ was constrained to take rational values due to the requirement of a crystallographic action, it can now be extended to any value. Upon summation over $g=0,1$, the partition function becomes
\begin{equation}
	\mathcal{Z} =  \tfrac{1}{2} \sum_{\tilde m,Q} \frac{\cos\left(\pi \epsilon_{+}(\tilde m_1+U_1\tilde m_2)\right) }{\sin\left(\pi \epsilon_{1}(\tilde m_1+U_1\tilde m_2)\right)  \, \sin\left(\pi \epsilon_{2}(\tilde m_1+U_1\tilde m_2)\right) } \, \frac{ [\Lambda_{2,10}(T,U,Y)]_{\tilde m,0,Q}}{\bar q}  \,.
\end{equation}
In the Nekrasov setup, $T^2$ factorises as the product of two circles with radii $R_1$, $R_2$ and we, henceforth, set $U_1=0$. With this choice
\begin{equation}
	[\Lambda_{2,10}(T,U,Y)]_{\tilde m,0,Q} \to \frac{R_1 R_2}{\tau_2} \,\bar q^{\frac{1}{2}Q\cdot Q} \, e^{-\frac{\pi}{\tau_2} (R_1^2 \, \tilde m_1^2 + R_2^2 \tilde m_2^2)} \, e^{2\pi i \tilde m_2 R_2 Y\cdot Q}  \,,
\end{equation}
and, upon Poisson-resummation over $\tilde m_2$, one may recast $\mathcal{Z}$ in the form
\begin{equation}
	\mathcal{Z} = \frac{R_1}{2\sqrt{\tau_2}} \sum_{\tilde m_1, m_2, Q} \frac{\cos (\pi\epsilon_{+}\tilde m_1)}{ \sin (\pi\epsilon_1\tilde m_1) \, \sin (\pi\epsilon_2 \tilde m_1)} \, e^{-\pi R_1^2 \tilde m_1^2/\tau_2 - \pi\tau_2 (Y\cdot Q- m_2/R_2)^2 } \, \bar q^{\frac{1}{2}Q\cdot Q -1 } \,.
	\label {FinalZN2}
\end{equation}
The $\text{SU}(2)$ symmetry enhancement point can be achieved, for instance, by taking $Q=\pm (1,-1,0^6)$. Therefore, in the 5d limit $R_2\to 0$, the free energy reads
\begin{equation}
	\begin{split}
	\mathcal{F} &\sim   R_1 \sum_{\tilde m_1 =1}^{\infty} \frac{-2\cos (\pi\epsilon_{+}\tilde m_1)}{ \sin (\pi\epsilon_1\tilde m_1) \, \sin (\pi\epsilon_2 \tilde m_1)}  \int_0^\infty \frac{dt}{t^{3/2}} \, e^{-\pi R_1^2 \tilde m_1^2/t - \pi t (Y_1-Y_2)^2 }  
		\\
		&= \sum_{n =1}^{\infty} \frac{1}{n}\, \frac{-2\cos (\pi n\epsilon_{+})}{ \sin (\pi n\epsilon_1) \, \sin (\pi n\epsilon_2)} \, e^{-2\pi n R_1 |Y_1-Y_2| }   \,,
	\end{split}
	\label{NekOk}
\end{equation}
which exactly reproduces the Nekrasov-Okounkov result in the case of two independent equivariant parameters $\epsilon_{1,2}$~\cite{Nekrasov:2003rj}.

\subsection{$\mathcal N=2^\star$ Heterotic String on the $\Omega$-Background}\label{N2star}

The previous construction can be readily extended to the case of $\mathcal{N}=2^\star$ gauge theories. In fact, $\mathcal{N}=2^\star$ corresponds to a spontaneous breaking of $\mathcal{N}=4$, and can be geometrically engineered \cite{Florakis:2015ied,Angelantonj:2017qeh,Samsonyan:2017xdi} via the Scherk-Schwarz mechanism \cite{Scherk:1978ta,Kounnas:1989dk,Kounnas:1988ye,Kiritsis:1997ca}. As a result, the $T^4 /\mathbb{Z}_2$ orbifold is now freely acting and, effectively, the only modification to the partition function \eqref{ZN2} is the replacement 
\begin{equation}
	\Gamma_{2,2}(T,U) \to \Gamma_{2,2}\big[ {\textstyle{h \atop g}}\big] (T,U) \,,
\end{equation}
where, in the Hamiltonian representation, the modified Narain lattice involves the phase $(-1)^{m_2 g}$ along with the shift $n_2\to n_2 + h/2$.
The sum over spin structures may be again performed in a similar fashion while, in the same field theory limit, the sum over $g=0,1$ now yields
\begin{equation}
	\sum_{g=0,1} (-1)^{m_2 g}\, \sin \left[ \tfrac{\pi}{2}(\tilde m_1\epsilon_{+}+g) \right]\, \sin \left[ \tfrac{\pi}{2}(\tilde m_1\epsilon_{+}-g) \right]= \sin^2  (\tfrac{\pi}{2}\tilde m_1\epsilon_{+}) -(-1)^{m_2} \cos^2  (\tfrac{\pi}{2}\tilde m_1\epsilon_{+}) \,.
\end{equation}
As a result, eq. \eqref{FinalZN2} takes the form
\begin{equation}
	\mathcal{Z} =   \frac{R_1}{2\sqrt{\tau_2}}\, \sum_{\tilde m_1, m_2,Q} \frac{ e^{- \pi\tau_2 (Y\cdot Q- \mu - 2m_2/R_2)^2} -\cos(\pi \epsilon_{+}\tilde m_1) \,e^{- \pi\tau_2 (Y\cdot Q - 2m_2/R_2)^2}}{\sin\left(\pi \epsilon_{1}\tilde m_1\right)  \, \sin\left(\pi \epsilon_{2}\tilde m_1\right) } \, e^{-\pi R_1^2 \tilde m_1^2/\tau_2  } \, \bar q^{\frac{1}{2}Q\cdot Q -1 }  \,,
\end{equation}
where $\mu = 1/R_2$ is the mass of the adjoint hypermultiplet. In the 5d fixed-$\mu$ limit, we match the perturbative contribution to the free energy 
\begin{equation}
	\mathcal{F} \sim \sum_{n =1}^{\infty} \frac{1}{n}\, \frac{2\,e^{-2\pi n R_1 |Y_1-Y_2-\mu |} -2\,\cos (\pi n\epsilon_{+})\, e^{-2\pi n R_1 |Y_1-Y_2|}}{ \sin (\pi n\epsilon_1) \, \sin (\pi n\epsilon_2)} 
\end{equation}
of $\mathcal{N}=2^\star$  $\text{SU}(2)$ gauge theory.


\section{The Refined Topological Amplitudes}\label{Section3}

In the previous section it was shown that the propagation of the heterotic string on Melvin spaces correctly reproduces the perturbative free energy of an $\mathcal{N}=2$ gauge theory on the $\Omega$-background. In the case $\epsilon_{+}=0$, Nekrasov and Okounkov conjectured \cite{Nekrasov:2002qd,Nekrasov:2003rj} that the full (non-perturbative) free energy be captured by the genus-$g$ partition function $\mathcal{F}_g$ of the type II topological string on a suitable CY space. Actually, the full $\mathcal{F}_g$ computes the exact scattering amplitude involving two gravitons and $2g-2$ graviphotons in the physical (non-topological) type II superstring on the genus-$g$ Riemann surface. These amplitudes may be alternatively computed in the dual heterotic theory. Since the heterotic dilaton lies in a vector multiplet, the amplitude receives perturbative contributions only at genus one, but further non-perturbative corrections are now present. Therefore, in the field theory limit and around a suitable gauge symmetry enhancement point, the heterotic genus-one amplitude captures the perturbative contribution to the Nekrasov free energy. 

As proposed in \cite{Nekrasov:2002qd,Nekrasov:2003rj}, it is natural to assume that a similar connection also persists in the general $\epsilon_{+}\neq 0$ case. According to the program initiated in \cite{Antoniadis:2010iq,Antoniadis:2013bja}, we expect that the new scattering amplitude should be extended by the introduction of additional vertex operators. To correctly identify this extension, we follow \cite{Angelantonj:2019qfw} and utilise the geometric realisation of the  $\Omega$-background as a Melvin space.

The prototype of a Melvin space \cite{Melvin:1963qx,Russo:1994cv} deforming a single plane is characterised by the metric
\begin{equation}
	ds^2 = G_{MN} dx^M dx^N= d\rho^2 + \rho^2 \left( d\phi +\epsilon dy\right)^2 + dy^2 + \ldots \,,
\end{equation}
where $\rho$ and the angle $\phi \in [0,2\pi)$ are the polar coordinates on $\mathbb{R}^2$, $y$ parametrises the unit circle $S^1$, and the ellipses stand for additional spectator coordinates. The Kaluza-Klein Ansatz 
\begin{equation}
	g_{\mu\nu} = G_{\mu\nu} - \frac{G_{\mu y}G_{\nu y}}{G_{yy}} \,,\qquad A_{\mu} = \frac{G_{\mu y}}{G_{yy}} \,, \qquad e^{2\sigma} = G_{yy} \,,
\end{equation} 
identifies the lower dimensional geometry
\begin{equation}
		ds^2 = d\rho^2 + \rho^2 \, F(\rho)\,d\phi^2 +\ldots \,, \qquad A = \epsilon\rho^2\,F(\rho)\,d\phi \,, \qquad e^{2\sigma} = F(\rho) \,,
\end{equation}
describing a magnetic flux-tube in a properly curved background, where $F^{-1}(\rho) = 1+\epsilon^2 \rho^2$. Notice that this same space can be equivalently described by a flat metric without magnetic flux-tubes, provided one introduces the new coordinate $\varphi = \phi +\epsilon y$ which, however, no longer describes an angle. Indeed, as $y$ winds $n$ times around $S^1$, the complex coordinate $z= \rho\,e^{i\varphi}$ is rotated by an angle $2\pi n\epsilon$. Therefore, the magnetic flux-tube is mapped to a rotation. This property of the Melvin space has an equivalent in the topological amplitudes of~\cite{Antoniadis:1995zn}. There the scattered fields are anti-self-dual graviphotons,  and the deformed $\sigma$-model for the corresponding generating function reads
\begin{equation}
	S_{\text{def}}(\epsilon) =  \int d^2 \sigma\,\chi\left( Z^1 \bar \partial Z^2 + \bar Z^2 \bar\partial \bar Z^1 \right) \,,
\end{equation}
which clearly involves the angular momentum generators, rotating the two planes of Euclidean spacetime, 
$(Z^1_R, Z^2_I)$ and $(Z^2_R, Z^1_I)$,
by opposite angles $\pm2\pi\chi$, where $Z^i=Z^i_R+i\,Z^i_I$. Following the steps of~\cite{Antoniadis:1995zn}, the complexified angle $\chi=\epsilon (\tilde{m}_1+U\tilde{m}_2+\tau(n_1+Un_2))$ is now dressed with the winding numbers around the complex worldsheet coordinate $X$ in $T^2$. This equivalence lies at the heart of the interpretation of the topological amplitudes as the string background given in \cite{Angelantonj:2019qfw}.

Using these relations, along with the geometry of the Melvin space discussed in Section \ref{Section2}, we identify the following structure for the deformed $\sigma$-model in the presence of two independent equivariant parameters
\begin{equation}
	S_{\text{def}}(\epsilon_{-},\epsilon_{+}) =  \int d^2 \sigma \left[ \chi_{-}  \left( Z^1 \bar \partial Z^2 + \bar Z^2 \bar\partial \bar Z^1 \right) +\chi_{+} \left( \bar Z^1 \bar\partial Z^2 + \bar Z^2\bar\partial Z^1 + \bar Z^3\bar\partial Z^4 + \bar Z^4\bar\partial Z^3 \right) \right] \,.
		\label{2Edef}
\end{equation}
As usual, the complex worldsheet coordinates $Z^1$ and $Z^2$ parametrise the Euclidean spacetime, while $Z^3$ and $Z^4$ parametrise the internal $T^4/\mathbb {Z}_2$, and $\chi_{\pm}$ are proportional to the angles $\epsilon_{\pm}$. 
By going to real coordinates as above, one can easily see that the two planes $(Z^1_R, Z^2_I)$ and $(Z^2_R, Z^1_I)$ are now rotated by two independent angles $2\pi\epsilon_1$ and $2\pi\epsilon_2$, while the planes $(Z^3_R, Z^4_I)$ and $(Z^4_R, Z^3_I)$ are rotated by $\pi\epsilon_+$ --- the latter corresponding to the $SU(2)_R$ rotation.
In the special case $\epsilon_{+}=0$, one recovers the generating function of~\cite{Antoniadis:1995zn}, which was shown in \cite{Angelantonj:2019qfw} to reproduce the free energy of the heterotic string on the Melvin background.  

The structure of \eqref{2Edef} leads to a generic scattering amplitude of the type
\begin{equation}
	\mathcal{A}_{g,n,m} = \langle \mathcal{V}_{\text{grav}}^2\, \mathcal{V}_{\text{gph}}^{2g-2}\, \mathcal{V}_{\text{gauge}}^{2n} \, \mathcal{V}_{\text{flux}}^{2m} \rangle \,,
	\label{Amplitude}
\end{equation}
which, aside from the two gravitons and the $2g-2$ anti-self-dual graviphotons of \cite{Antoniadis:1995zn}, now involves $2n$ self-dual gauge fields in the dilaton vector multiplet, together with $2m$ insertions of self-dual magnetic fluxes along the $T^4/\mathbb{Z}_2$. Although the $\mathbb{Z}_2$ projection removes all gauge fields with index along the orbifold space, internal fluxes survive the projection and can participate in the scattering. These fluxes actually realise the R-symmetry rotation required by  supersymmetry. In eq. \eqref{Amplitude},  $\mathcal{V} = \int d^2 z\,V(z,\bar z;p)$ where the $V(z,\bar z;p)$'s are the vertex operators in the 0 ghost picture carrying momentum $p$. In what follows, we suppress the explicit dependence on the position unless necessary. Adopting the specific kinematic configuration of \cite{Antoniadis:1995zn, Antoniadis:2010iq}, and stripped off of their physical polarisation tensors, the vertex operators of eq. \eqref{Amplitude} read
\begin{equation}
	\begin{split}
		 V_{\rm grav}(p) &= (\partial Z^2 - ip\,\psi^1\psi^2)\,\bar\partial Z^2\, e^{ipZ^1} \,, \\
	 	 V_{\rm gph}(p) &= (\partial X - i p\, \psi^1 \Psi)\, \bar\partial Z^2 \, e^{ipZ^1}  \,, \\
		 V_{\rm gauge}(p) &= (\partial X- i p\, \psi^1 \Psi)\, \bar\partial \bar Z^2\, e^{i  p  Z^1} \,, \\
		 V_{\rm flux}(P) &= (\partial X - iP\,\psi^3 \Psi)\, \bar\partial\bar Z^4\, e^{iP Z^3}  \\	
		                         &\rightarrow (\partial X - iP\,\psi^3 \Psi)\, iP Z^3 \,\bar\partial\bar Z^4\,, \\
	\end{split}
		\qquad
	\begin{split}
		 V_{\rm grav}(\bar p) &= (\partial \bar Z^2 - i \bar p\, \bar\psi^1\bar\psi^2)\,\bar\partial \bar Z^2\, e^{i\bar p\bar Z^1}  \,, \\
		 V_{\rm gph}(\bar p) &= (\partial X - i\bar p\, \bar \psi^1 \Psi)\, \bar\partial \bar Z^2 \, e^{i\bar p \bar Z^1}  \,, \\
		 V_{\rm gauge}(\bar p) &= (\partial X- i\bar p\, \bar\psi^1 \Psi)\, \bar\partial Z^2\, e^{i \bar p\bar Z^1} \,, \\
		 V_{\rm flux}(\bar P) &= (\partial X - i\bar P\,\bar\psi^3 \Psi)\, \bar\partial  Z^4\, e^{i\bar P \bar Z^3}  \\
		                                &\rightarrow (\partial X - i\bar P\,\bar\psi^3 \Psi)\, i\bar P \bar Z^3\,\bar\partial  Z^4\,. \\
	\end{split}
\end{equation}
In these expressions,  the complex $\psi^i$ and $\Psi$ are the worldsheet fermionic superpartners of $Z^i$ and $X$, respectively. The complex momenta $p$ run over the Euclidean spacetime directions, while $P$ refers to the $\bar Z^3$ direction of K3. 
Notice that, although $V_{\rm flux}$ does not propagate physical degrees of freedom, the terms $PZ^3\bar\partial \bar{Z}^4$ and $\bar P \bar Z^3 \bar\partial Z^4$ in its expansion represent a quantised flux, invariant under the orbifold action. 

The only non-vanishing contributions to the amplitude involving the two gravitons arise from the fermionic contractions, which soak up four fermionic zero modes. Among the remaining vertex operators, there are no contractions involving $\partial X$ or $\Psi$. As a result, the contributions of their quantum parts to the functional integral cancel that of the $(b,c)$ and $(\beta,\gamma)$ (super)ghost systems, so that $\partial X$ contributes only zero modes.

The term in the string effective action we are interested in, involves two anti-self-dual Riemann tensors and a number of (anti) self-dual field strengths, and can be computed from $\mathcal{A}_{g,n,m}$ by extracting the appropriate powers of momenta
\begin{equation}
	\mathcal{A}_{g,n,m} \supset \left[ (p\bar p)^2 \prod_{i=1}^{g-1} (p\bar p)_i \prod_{k=1}^{n} (p\bar p)_k \prod_{\ell=1}^{m} (P\bar P)_{\ell} \right]\, \mathcal{C}_{g,n,m} \,,
\end{equation}
with our choice of kinematics. $\mathcal{C}_{g,n,m}$ now involves only bosonic correlators, and reads
\begin{equation}
	\begin{split}
	\mathcal{C}_{g,n,m} &=  \Big\langle \,\prod_{i=1}^{g} \int d^2 x_i\, Z^1\bar\partial Z^2(x_i) \, \int d^2 y_i\, \bar Z^1\bar\partial \bar Z^2(y_i)\, \prod_{k=1}^{n} \int d^2 u_k\, \bar Z^1 \bar\partial Z^2(u_k)\, \int d^2 v_k\, \bar Z^2\bar\partial Z^1(v_k) \, \\ 
		&\qquad \times \prod_{\ell=1}^{m} \int d^2 r_\ell \, \bar Z^3 \bar\partial Z^4(r_\ell)\, \int d^2 s_{\ell} \, \bar Z^4\bar\partial Z^3(s_\ell) \,\Big\rangle \,.
	\end{split}
\end{equation}
These correlators may be conveniently computed by constructing the generating function 
\begin{equation}
	\mathcal{F}(\lambda_1,\lambda_2,\lambda_3) = \sum_{g,n,m=0}^{\infty} \frac{\lambda_1^{2g}\,\lambda_2^{2n}\, \lambda_3^{2m} }{(g!\,n!\,m!)^2} \,\mathcal{C}_{g,n,m}\,,
	\label{RefGF}
\end{equation}
thus, generalising the approach in \cite{Antoniadis:1995zn}.
The structure of the correlators implies that the r.h.s. can be exponentiated into a new $\sigma$-model
\begin{equation}
	\mathcal{F}(\lambda_1,\lambda_2,\lambda_3) = \big\langle \, e^{ -  \int d^2 \sigma \left[ \lambda_{1}  \left( Z^1 \bar \partial Z^2 + \bar Z^2 \bar\partial \bar Z^1 \right) +\lambda_{2} \left( \bar Z^1 \bar\partial Z^2 + \bar Z^2\bar\partial Z^1\right) +\lambda_3\left( \bar Z^3\bar\partial Z^4 + \bar Z^4\bar\partial Z^3 \right) \right]   } \, \big\rangle \,,
		\label{Deformation}
\end{equation}
which reduces to the deformed action \eqref{2Edef}, involving the complexified Melvin angles, upon the identification $\lambda_1 \to \chi_{-}$ and $\lambda_{2,3} \to \chi_{+}$.

 Note that, in writing \eqref{Deformation}, the functional integral is evaluated with respect to the original free action for the bosonic fields, $S_0=\frac{1}{\pi}\int d^2\sigma \sum_{i=1}^{4}(\partial Z^i\bar\partial \bar Z^i+\partial\bar Z^i\bar\partial Z^i)$. The functional integral \eqref{Deformation} is still Gaussian and is, thus, amenable to an exact evaluation. However, some care is needed when dealing with zero modes of the compact K3 coordinates. To evaluate \eqref{Deformation}, we decompose the fields into their classical part and their quantum fluctuations, $Z^i=Z_{\rm classical}^i + Z_{\rm quantum}^i$. The quantum part is independent of the compactness or not of the $Z$-coordinate, and can be computed via zeta-function regularisation techniques, as in \cite{Antoniadis:1995zn}. For the classical part, compact and non-compact coordinates yield instead different contributions. Focusing on the compact $Z^3, Z^4$ coordinates, the deformed equations of motion
\begin{equation}
	\partial\bar\partial Z^3 - \pi\lambda_3\, \bar\partial Z^4 =0 \,, \qquad \partial\bar\partial Z^4 - \pi\lambda_3 \,\bar\partial Z^3 =0\,,
\end{equation}
can be conveniently diagonalised
\begin{equation}
	\partial \bar\partial Z^{\pm} \mp \pi\lambda_3 \,\bar\partial Z^{\pm} = e^{\pm\pi\lambda_3 z}\partial\bar\partial \left( e^{\mp\pi\lambda_3 z} Z^{\pm}\right) =0\,,
\end{equation}
 in the $Z^{\pm} = Z^3\pm Z^4$ basis, with $z$ being the complex coordinate on the worldsheet torus. This implies that the combination $\hat {Z}^{\pm} = e^{\mp\pi\lambda_3 z}\, Z^{\pm}$ obeys the undeformed d'Alembert equation, although with the modified boundary condition
\begin{equation}
	\hat Z^{\pm}(\sigma_1+1,\sigma_2) = e^{\mp i\pi\lambda_3}\, \hat{Z}^{\pm}(\sigma_1,\sigma_2) \,,
\end{equation}
and similarly for the periodicity of $\sigma_2$. It is then clear that, by Lefschetz fixed-point theorem, the zero mode associated to these compact coordinates contributes $16\sin^4 (\pi\lambda_3/2)$ to the functional integral. 

Taking this into account, along with the fact that the $\lambda_i$ rotation angles are actually dressed by the zero modes of $X$, the full result reads
\begin{equation}
		\mathcal{F}(\lambda_1,\lambda_2,\lambda_3)  \sim  \int_{\mathcal F}\frac{d^2\tau}{\tau_2}\sum_{h,g=0,1}{\sum_{\tilde m,n,Q}}'  \,\frac{\zeta\big[ {\textstyle{ h\atop g}} \big](\tfrac{\lambda_3}{2})\,e^{-\frac{\pi}{2\tau_2}(\bar\xi_1^2+\bar\xi_2^2+\frac{1}{2}\bar\xi_{3}^2)} }{\bar\eta^{4}\, \bar\vartheta_1(\xi_1)\, \bar\vartheta_1(\xi_2)\,\bar\theta\big[ {\textstyle{ 1+h\atop 1+g}} \big](\frac{\xi_{3}}{2})\, \bar\theta\big[ {\textstyle{ 1-h\atop 1-g}} \big](\frac{\xi_{3}}{2})  } \,
		  \,[\Lambda_{2,10}(T,U,Y)]_{\tilde m,n,Q} \, \bar \Gamma_8 \big[ {\textstyle{ h\atop g}} \big] \,,
\label{BPSfreeL}
\end{equation}
where $\xi_i = \lambda_i  (\tilde{m}_1+U\tilde{m}_2+\tau(n_1+Un_2))$. 
Hence, the generating function of our amplitude precisely matches the partition function of the heterotic string on the Melvin background \eqref{BPSfree}, upon the identification $\lambda_1\to\epsilon_{-}$, $\lambda_{2,3}\to\epsilon_{+}$, and $\xi_i \to \chi_i$.

The same analysis may of course be repeated in the case of an $\mathcal N=2^\star$ theory, the only modification being the free action of the K3-realising orbifold, as in Section \ref{N2star}.


\section{Refined Couplings}\label{Section4}

In was shown in \cite{Antoniadis:1995zn} that the higher-derivative F-terms
\begin{equation}
	\int d^4\theta\,\mathcal F_g(X)\,(W^{ij}_{\mu\nu}W^{\mu\nu}_{\ij})^{g} = \mathcal F_g(\phi)\,R_{(-)\mu\nu\rho\sigma}R_{(-)}^{\mu\nu\rho\sigma}\,(F^{G}_{(-)\lambda\tau}F_{(-)}^{G\; \lambda\tau})^{g-1} + \ldots
\end{equation}
are related to topological amplitudes involving two anti-self-dual Riemann tensors $R_{(-)}$ and $2g-2$ anti-self-dual graviphoton field strengths $F^{G}_{(-)}$.
Here, 
 \begin{equation}
 	W_{\mu\nu}^{ij} = F^{G,ij}_{(-)\mu\nu} + \theta^{[i}B^{j]}_{(-)\mu\nu} +(\theta^i \sigma^{\rho\lambda} \theta^j) R_{(-)\mu\nu\rho\lambda}+\ldots
	\label{Weyl}
 \end{equation}
is the Weyl superfield, whose lowest component is the anti-self-dual graviphoton, and
\begin{equation}
	X^{I} = \phi^I + \theta^i\lambda_i^I+\tfrac{1}{2}F^I_{(-)\mu\nu}\epsilon_{ij} (\theta^i \sigma^{\mu\nu} \theta^j) + \ldots 
\end{equation}
is an $\mathcal N=2$ chiral superfield, associated to the $I$-th vector multiplet. The couplings $\mathcal F_{g}$ are precisely the coefficients of the $\epsilon$-expansion of the generating function for the topological amplitudes of \cite{Antoniadis:1995zn} and, in the field theory limit, coincide with the coefficients in the expansion 
\begin{equation}
	\tilde{\mathcal F}(\epsilon) = \sum_{g} \tilde{\mathcal F}_g \, \epsilon^{2g} \,,
\end{equation}
of the Nekrasov free energy $\tilde{\mathcal F}(\epsilon)$ for the case of a single equivariant parameter $\epsilon$. Clearly,  $\epsilon$ is associated to the anti-self-dual field strength of the graviphoton.

Following \cite{Antoniadis:1995zn}, one could ask which terms in the effective action are computed by the amplitudes \eqref{Amplitude} considered in the present work, and how they connect to the Nekrasov free energy $\tilde{\mathcal F}(\epsilon_{-},\epsilon_{+})$ in the case where both equivariant parameters $\epsilon_{\pm}$ are turned on. The structure of the amplitude \eqref{Amplitude} spells out the presence of the higher-derivative terms
\begin{equation}
	\mathcal F_{g,n,m}(\phi)\,R_{(-)}^2\,(F^{G}_{(-)})^{2g-2}\, (F^{S}_{(+)})^{2n} (M_{(+)})^{2m} \,,
	\label{HigherDeriv}
\end{equation}
where $F^S_{(+)}$ is the self-dual field strength associated to the dilaton vector multiplet and $M_{(+)}$ is a self-dual flux through the K3, while $\phi$ collectively denotes the moduli scalars of the vector multiplets. 

The couplings $\mathcal F_{g,n,m}$ are computed by expanding the generating function \eqref{BPSfreeL}, 
\begin{equation}
	\mathcal F(\lambda_1,\lambda_2,\lambda_3) = \sum_{g,n,m} \mathcal{F}_{g,n,m} \,\lambda_1^{2g} \lambda_2^{2n} \lambda_3^{2m}\,,
\end{equation}
where we associate $\lambda_1$ with the anti-self-dual field strength of the graviphoton, $\lambda_2$ with the self-dual field strength $F^S$ and $\lambda_3$ with the internal self-dual flux $M$.  In the field theory limit, we obtain
\begin{equation}
	\mathcal F(\lambda_1,\lambda_2,\lambda_3) \sim \sum_{k=1}^{\infty} \frac{1}{k} \, \frac{-2\cos(\pi k\lambda_3)}{\sin [\frac{\pi}{2} k(\lambda_2+\lambda_1)]\, \sin [\frac{\pi}{2} k(\lambda_2-\lambda_1)]}\,e^{-ka}\,,
\end{equation}
where $a=2\pi R_1 |Y_1-Y_2|$ is the $U(1)$ modulus parametrising the distance from the $SU(2)$ enhanced symmetry point.

The connection of eq. \eqref{RefGF} with the refined Nekrasov free energy
\begin{equation}
	 \tilde{\mathcal F}(\epsilon_{-},\epsilon_{+}) = \mathcal F(\epsilon_{-},\epsilon_{+},\epsilon_{+})  \,,
\end{equation}
follows from the identification of the field strengths and fluxes with the two equivariant parameters, $\lambda_1=\epsilon_{-}$ and $\lambda_{2}=\lambda_{3}=\epsilon_{+}$.
Matching the coefficients of the homologous powers of $\epsilon_{\pm}$, we obtain the desired relation 
\begin{equation}
	\tilde{\mathcal F}_{g,N} = \sum_{k=0}^{N} \mathcal F_{g,k,N-k} \,.
	\label{RTANO}
\end{equation}
Clearly, setting $N=0$ in \eqref{RTANO} yields the unrefined topological amplitude of \cite{Antoniadis:1995zn}, in accordance with our interpretation of $\mathcal F(\lambda_1,\lambda_2,\lambda_3)$ as an extension of $\mathcal F(\epsilon)$. 

Note that the presence of the fluxes $M$ is instrumental in recovering the Nekrasov result with both equivariant parameters turned on. Indeed, in the absence of the fluxes, the corresponding amplitudes $\mathcal C_{g,n,0}$, first computed in \cite{Antoniadis:2010iq}, fail to reproduce the signature of vector multiplets, \emph{i.e.} they miss the cosinus factor in the numerator of eq. \eqref{NekOk}. 

The higher-derivative terms \eqref{HigherDeriv} can be seen to arise from an integral over half the superspace
\begin{equation}
\label{EFTcouplings}
  	\int d^4\theta~  \sum_{k=0}^N {\mathcal F}_{g,k,N-k}(X^I)\,  W^{2g}\, \Upsilon_S^{k} \, M_{(+)}^{2N-2k} \,,
\end{equation}
which generalises the effective action of \cite{Antoniadis:2010iq,Antoniadis:2013bja} by the presence of the self-dual flux.
Here, $\Upsilon_S$ encodes the contribution of the self-dual field strengths $F^S_{(+)}$ in the dilaton vector multiplet. In the simplest case\footnote{In principle, one could replace the dilaton superfield in~\eqref{Upsilon} by a function $h({\hat X}^I,({\hat X}^I)^\dagger)$, introducing even a (non-holomorphic) dependence on additional moduli, that should be taken into account in the definition of the coefficients of the series~\eqref{EFTcouplings}. Extracting the exact effective supergravity description then requires a more detailed analysis of the string amplitudes involved.}, this superfield is built as the chiral projection \cite{Antoniadis:2010iq}
\begin{equation}
\label{Upsilon}
	\Upsilon_S = \tfrac{1}{2}\,\Pi \left(\frac{({\hat X}^S)^\dagger}{X^0}\right)^2=  \,\bigr(F^S_{(+)}\bigr)^2+\dots\,,
\end{equation}
where $\hat X^S = X^S/X^0$ is the dilaton superfield, and $X^0$ is the compensator superfield of $\mathcal N=2$ conformal supergravity~\cite{deRoo:1980mm}. Here, the projection operator $\Pi=(\epsilon_{ij}\bar D^i \bar\sigma_{\mu\nu} \bar D^j)^2$ is built out of spinor supercovariant derivatives of the $\mathcal N=2$ superconformal algebra, with the action
\begin{equation}
	\Pi\, X =0 \,, \qquad \Pi\, X^\dagger = 96\,\Box X \,,	
\end{equation}
on a generic chiral vector superfield $X$.
Since $M_{(+)}$ is not dynamical but a background flux, the couplings $\mathcal A_{g,k}$ of the higher-derivative terms
\begin{equation}
	\mathcal A_{g,k}(\phi)\,R_{(-)}^2\,(F^{G}_{(-)})^{2g-2}\, (F^{S}_{(+)})^{2k} \,,
\end{equation}
in the effective action involving two anti-self-dual Riemann tensors, $2g-2$ anti-self-dual graviphoton field strengths and $2k$ self-dual dilaton vector field strengths in the flux background, are given by
\begin{equation}
	{\mathcal A}_{g,k} = \sum_{N\ge k}^{} \mathcal F_{g,k,N-k}M_{(+)}^{2N-2k} \,.
\end{equation}


\section{Conclusions}\label{SectionConclusions}

In this work we have identified the refinement of the topological amplitudes \cite{Antoniadis:1993ze}, in the heterotic weak coupling limit, associated to the full $\Omega$-background. By generalising the results of \cite{Angelantonj:2019qfw} to include also the second equivariant parameter $\epsilon_{+}$, we obtained an exact realisation of the $\Omega$-background in string theory as a 10d Melvin space, where two independent rotations in the Euclidean 4d spacetime are now accompanied by an appropriate supersymmetry-preserving rotation inside K3, together with a shift along a one-cycle of $T^2$. 
The upshot of our analysis is the fact that the generalised Melvin space corresponds to an exactly solvable world-sheet $\sigma$-model.
Working in the $T^4/\mathbb Z_2$ orbifold limit of K3, we computed the one-loop heterotic vacuum amplitude and showed that, expanded around a point of enhanced SU(2) gauge symmetry, the field theory limit of our result correctly reproduces the perturbative part of the Nekrasov free energy with both deformation parameters turned on.

By requiring that the refined amplitudes exponentiate to the very same deformed $\sigma$-model as the one realising our 10d Melvin geometry, we were able to explicitly identify their vertex operators. We found that, besides the two anti-self-dual Riemann tensors and a number of anti-self-dual graviphoton field strengths, the additional insertions correspond to self-dual field strengths of the dilaton multiplet, as well as to self-dual magnetic fluxes along K3. We explicitly evaluated the generating function of the refined amplitudes and confirmed that it indeed matches the vacuum amplitude in our Melvin background, upon identifying the anti-self-dual field strengths with $\epsilon_{-}$ and the additional self-dual field strengths with $\epsilon_{+}$. Importantly, our refined amplitudes receive contributions only from BPS states and their generating function correctly reduces to the ``unrefined'' case upon setting $\epsilon_{+}=0$.
A further non-trivial check of our approach is the fact that hypermultiplet moduli manifestly decouple both at the level of the generating function, as well as of the partition function on our 10d Melvin background. As a result, our refined amplitudes, stemming from the geometrical engineering of the full $\Omega$-background in string theory, provide a well-defined extension of the topological amplitudes of \cite{Antoniadis:1993ze}, that we expect could serve as a starting point for a worldsheet definition of the refined topological string.

Certain interesting open questions remain. Although the fields associated to the vertex operators involved in the refined amplitudes were identified, the superfield description of the corresponding higher-derivative terms in the effective action is not straightforward. This is because the chiral projection discussed in Section \ref{Section4} introduces potential ambiguities, associated to the possibility of including a general function of the dilaton superfield (and even of all vector moduli). This may be related to the problem of extracting the form of the holomorphic anomaly equation at the level of the partition function and the generated amplitudes, which constitutes by itself a second open question. Another important problem is the extension of our analysis to the non-perturbative level, which could in principle be studied on the type II side, since the latter provides the exact answer in the one-parameter case of vanishing $\epsilon_+$. Since the Melvin space corresponds to a freely-acting orbifold, it should in principle be realised also on the dual type IIA superstring compactified on an appropriate CY manifold.

\section*{Acknowledgements}
C.A. would like to thank LPTHE Paris, the CERN Theory Division and the Institute for Theoretical Physics of the University of Bern for their warm hospitality during several stages of this work. The work of C.A. is partially supported by the MIUR-PRIN contract 2017CC72MK-003. 
 The work of H.J.    is supported by the Royal Society grant, ``Relations, Transformations, and
Emergence in Quantum Field Theory'', and by the Science and Technology Facilities Council (STFC) Consolidated Grant ST/T000686/1 ``Amplitudes, strings \& duality''.

\bibliographystyle{unsrt}

\end{document}